\documentclass[12pt]{iopart}
\usepackage{graphicx,color}
\usepackage{verbatim}
\usepackage{xspace}
\newcommand{\targetwave}{$\psi_{\mathrm{G}}$\xspace}
\newcommand{\trialwave}{$\Psi_{\mathrm{t}}$\xspace} 
\newcommand{\zeff}{$Z_{\mathrm{eff}}$\xspace}
\newcommand{\ehmol}{$\left(e^{+}-\mathrm{H}_{2}\right)$\xspace}
\newcommand{\hmol}{$\mathrm{H}_{2}$\xspace}
\newcommand{\ps}{$\eta_{\mathrm{v}}$\xspace}
\newcommand{\hyl}{Hylleraas-type\xspace}
\newcommand{\pc}{$P_{\mathrm{C}}$\xspace}
\newcommand{\targA}{$\psi_{\mathrm{G}}^{(\mathrm{A})}$\xspace}
\newcommand{\targB}{$\psi_{\mathrm{G}}^{(\mathrm{B})}$\xspace}
\definecolor{myRed}{rgb}{0.9,0,0}
\definecolor{myGreen}{rgb}{0,0.4,0}

\begin{document}
 
\title[Accurate target wave functions in \ehmol scattering]{The importance of an accurate target wave function in variational calculations for \ehmol scattering}

\author{J N Cooper$^1$, E A G Armour$^1$ and 
M Plummer$^2$}

\address{$^1$ School of Mathematical Sciences, University Park, Nottingham NG7 2RD, UK}
\address{$^2$ STFC Daresbury Laboratory, Daresbury, Warrington, Cheshire WA4 4AD, UK}
\ead{james.cooper@maths.nottingham.ac.uk}

\begin{abstract}
Using the complex Kohn method, we have calculated variational values of phase shifts and the annihilation parameter, \zeff, for the elastic scattering of positrons by molecular hydrogen. Our results are sensitive to small changes in the accuracy of the wave function representing the target hydrogen molecule. We have developed a systematic approach to demonstrate that, at low positron energies, there are particular forms of the Kohn trial wave function for which the results of variational calculations are not reliable, even when the target wave function accounts for as much as $96.8 \%$ of the correlation energy of \hmol. We find that reliable results can be recovered if our calculations are extended to admit more sophisticated target wave functions accounting for $99.7 \%$ of the correlation energy. Remaining discrepancies between theory and experiment are briefly discussed.
\end{abstract}
\submitto{\jpb}
\maketitle

\section{Introduction}\label{s:introduction}

In a previous article \cite{Cooper2007}, preliminary calculations of phase shifts were presented for the elastic scattering of low energy positrons by molecular hydrogen. The calculations used a generalized form of the Kohn variational method \cite{Kohn1948} and were carried out for the lowest partial wave of $\Sigma_{\mathrm{g}}^{+}$ symmetry, which has been shown \cite{Armour1990} to be the only significant contributor to scattering processes for incident positron energies below 2 eV. The Kohn calculations were the first for the \ehmol system to treat the potential term corresponding to the \hmol molecule explicitly. This introduced a complication \cite{Cooper2007} that had been avoided in earlier applications \cite{Armour1990,Armour1987,Armour1988} of the Kohn method by the use of the method of models \cite{Drachman1972}. 

Our implementation of the Kohn method has recently been made considerably more accurate. This is due most notably to the inclusion in the trial wave function of a greater number of terms describing short-range correlations between the molecular electrons and the positron, as well as to the use of improved methods for eliminating numerical anomalies due to so-called Schwartz singularities \cite{Schwartz1961,Nesbet1968}. We intend to publish details of our analysis of Schwartz singularities in a forthcoming article. In this paper, we will discuss results that have arisen during our efforts to improve the quality of the Kohn calculations by introducing a more flexible description of leptonic correlations than has previously been used.

Our earlier calculations \cite{Cooper2007} included in the description of the electron-positron correlation a number of terms that were linear in the electron-positron distance. The importance of such terms was first demonstrated by Hylleraas \cite{Hylleraas1929} and their inclusion greatly increases the speed at which the description of the electron-positron interaction converges. As explained by Armour \cite{Armour1985}, this is due to the role of linear terms in the electron-positron distance in satisfying the Kato cusp condition \cite{Kato1957}. The use in the earlier Kohn calculations of \hyl functions in the electron-positron coordinate was seen to increase significantly both the variational estimate, \ps, of the phase shift and the estimate of \zeff, the positron annihilation parameter. The value of \zeff measures the correlation between the positron and the target molecule and can be regarded as the effective number of molecular electrons available for annihilation with the positron. Methods for determining theoretical values of \zeff in the context of our Kohn calculations have already been outlined \cite{Cooper2007}.

Discrepancies remained between the results of the earlier calculations and available experimental data, and our efforts to improve the theoretical model are ongoing. We have recently extended our Kohn calculations to assess for the first time the contribution to \ps and \zeff made by \hyl functions that are linear in the interelectronic distance. We have found that, when \hyl functions of this form are included in the description of the short-range leptonic correlations, the results of the Kohn calculations can be unreliable unless the wave function used to represent the target \hmol molecule is itself made more accurate by the inclusion of \hyl functions in the interelectronic distance. Our observations are similar to those made by Van Reeth and Humberston \cite{VanReeth1995} in their Kohn calculations of s-wave phase shifts for positron scattering by atomic helium. Close-coupling calculations of electron scattering using inexact target states \cite{Plummer2004,Zatsarinny2006} have also shown that a balance must be maintained between the complexity of the treatment of target and scattering function electronic correlation. 

Our findings are important as the calculations that we will show to be unreliable can involve target states that account for nearly $97 \%$ of the correlation energy \cite{BransdenJoachain2003} and, coincidentally, give agreement with experimental results. When efforts are made to make the calculation more accurate, this agreement is no longer observed and the effect of including \hyl correlation functions in the interelectronic distance is found to be small. To cope with these difficulties, we present a numerical technique to analyze the variational results of calculations using inexact target states. This technique allows us to distinguish between reliable calculations and calculations which are unreliable due to an imbalance of target and scattering electronic correlation.

\section{Theory}
\subsection{The complex Kohn variational method}

A description of the Kohn variational method is given elsewhere \cite{Kohn1948}. Calculations specific to the elastic scattering of positrons by molecular hydrogen have been carried out by Armour and coworkers \cite{Armour1990,Armour1987,Armour1988} and, more recently, by Cooper and Armour \cite{Cooper2007}. Only the essential aspects of the method will be repeated here.

The objective of the calculation is to obtain an accurate approximation to the exact scattering wave function, $\Psi$, from which it is straightforward to calculate variational estimates, \ps, of the phase shift, as well as estimates of \zeff. Approximate wave functions can be obtained by prescribing a trial wave function, \trialwave, whose general form is known but which depends linearly on $n$ unknown parameters. The basis of the Kohn method is the application of a stationary principle that allows optimal values of these parameters to be determined by solving a corresponding system of linear equations in the $n$ unknowns.

The complex Kohn method \cite{McCurdy1987} is an extension of the original variational approach and allows the trial wave function, \trialwave, to be complex-valued. It has been shown \cite{Schneider1988} that the use of complex trial functions can mitigate the effects of anomalous results due to Schwartz singularities that arise when the linear system of Kohn equations is numerically ill-conditioned. Our own investigations of Schwartz singularities, to be discussed in a future article, have confirmed that the results presented here are free of this type of anomalous behaviour.

In our calculations on \ehmol scattering, we have found it convenient to fix the nuclei at the equilibrium internuclear separation, $R=1.4$ a.u. and to label the electrons as particles $1$ and $2$, taking the positron to be particle $3$ and describing each particle by prolate spheroidal coordinates \cite{Flammer1957} $(\lambda_{j},\mu_{j},\phi_{j})$, $j\in\{1,2,3\}$. We have used a complex trial wave function of the form 

\begin{equation}\label{eq:complextrialwave}
\Psi_{\mathrm{t}} = \left(S + a_{\mathrm{t}}T + p_{0}\chi_{0}\right)\psi_{\mathrm{G}} + \sum_{i=1}^{M} p_{i}\chi_{i},
\end{equation}
where

\begin{equation}\label{eq:hankel}
 T=S+\rmi C,
\end{equation}

\begin{equation}\label{eq:SinOpenChannel}
S=\frac{N}{\lambda_{3} - 1}\sin\left[ c\left( \lambda_{3} - 1\right) \right]
\end{equation}
and

\begin{equation}\label{eq:CosOpenChannel}
 C = \frac{N}{\lambda_{3} - 1}\cos\left[c\left( \lambda_{3}-1\right) \right] \lbrace 1-\exp\left[-\gamma\left( \lambda_{3}-1\right)\right]\rbrace.
\end{equation}

The functions $S$ and $C$ are the same as those used in our previous Kohn calculations \cite{Cooper2007} and represent, respectively, the incident and scattered positrons asymptotically far from the target \hmol molecule. The shielding parameter, $\gamma$, ensures that $C$ is regular at the origin and, as before \cite{Cooper2007}, is taken to have the value $\gamma=0.75$. The constant $c$ is defined to be $c=kR/2$, $k$ being the magnitude of the positron momentum in atomic units. $N$ is a normalization constant. The unknowns $a_{\mathrm{t}}$ and $\{p_{0},...,p_{M}\}$ are the complex-valued constants to be determined by the Kohn variational method. The function \targetwave is an approximation to the ground state wave function of the unperturbed hydrogen molecule and is determined by the Rayleigh-Ritz variational method \cite{BransdenJoachain2003}. The general form of \targetwave will be discussed in section \ref{ss:thehydrogenmolecule}.

The short-range correlation functions, $\Omega=\{\chi_{0},...,\chi_{M}\}$, allow for the description of direct electron-positron and electron-electron interactions. $\chi_{0}$ is the same correlation function used in our previous calculation \cite{Cooper2007} and was introduced first by Massey and Ridley \cite{MasseyRidley1956}. The general form of the remaining functions, for ($1\leq i\leq M$), is

\begin{eqnarray}
\nonumber\chi_{i} &=& N\left[\lambda_{1}^{a_{i}}\lambda_{2}^{b_{i}}\mu_{1}^{c_{i}}\mu_{2}^{d_{i}}s_{13}(\theta_{i})+\lambda_{1}^{b_{i}}\lambda_{2}^{a_{i}}\mu_{1}^{d_{i}}\mu_{2}^{c_{i}}s_{23}(\theta_{i})\right]\\
\label{eq:SRCF}&\times&\lambda_{3}^{r_{i}}\mu_{3}^{s_{i}}s_{12}(\theta_{i})\exp\left[-\beta\left(\lambda_{1}+\lambda_{2}\right)-\alpha\lambda_{3}\right],
\end{eqnarray}
for prescribed basis states, $\{a_{i},b_{i},c_{i},d_{i},r_{i},s_{i},\theta_{i}\}$, comprising non-negative integers. The interparticle functions, $s_{pq}(\theta_{i})$, have the form

\begin{eqnarray}\label{eq:s12}
    s_{12}(\theta_{i}) = \left\{ \begin{array}{ll}
                     \rho_{12}=\frac{2}{R}r_{12}    & \quad (\theta_{i} = 1) \\
                     M_{12} \cos (\phi_1 - \phi_2)   & \quad (\theta_{i} = 2) \\
             1                 & \quad (\mathrm{otherwise}),
                     \end{array} \right.
\end{eqnarray}\\and
\begin{eqnarray}
    s_{j3}(\theta_{i}) = \left\{ \begin{array}{ll}
                     \rho_{j3}=\frac{2}{R}r_{j3}    & \quad (\theta_{i} = 3) \\
                     M_{j3} \cos (\phi_j - \phi_3)   & \quad (\theta_{i} = 4) \\
             1                 & \quad (\mathrm{otherwise}),
                     \end{array} \right.
\end{eqnarray}
for $j \in \{1,2\}$, where $r_{pq}$ is the distance between leptons $p$ and $q$. The inclusion of terms of the form $M_{pq}\cos(\phi_p - \phi_q)$, where

\begin{equation}\label{eq:mab}
 M_{pq}=\left[(\lambda_{p}^2-1)(1-\mu_{p}^2)(\lambda_{q}^2-1)(1-\mu_{q}^2)\right]^{1/2},
\end{equation}
is equivalent to considering terms in $r_{pq}^2$. The choice of the nonlinear parameters, $\alpha$ and $\beta$, will be discussed in section \ref{ss:opt}.

We have carried out Kohn calculations using two different sets of correlation functions, which for convenience we shall denote by $\Omega^{(1)}$ and $\Omega^{(2)}$. The set $\Omega^{(1)}$ has $M=279$ and, in addition to $\chi_{0}$, contains three subsets of 87 basis functions corresponding to $\theta_{i}=0$, $\theta_{i}=2$ and $\theta_{i}=4$, as well as 18 \hyl basis functions in the electron-positron coordinates, for which $\theta_{i}=3$. $\Omega^{(1)}$ has the same general form as the set of 99 correlation functions used in our earlier calculation \cite{Cooper2007}. The set $\Omega^{(2)}$ has $M=297$ and is identical to $\Omega^{(1)}$ but for the inclusion of a further 18 \hyl basis functions in the interelectronic coordinate, for which $\theta_{i}=1$. Further details of the individual basis functions used are available from the corresponding author.

\subsection{The hydrogen molecule}\label{ss:thehydrogenmolecule}

Although the Schr\"{o}dinger equation for the hydrogen molecule cannot be solved exactly, very accurate numerical approximations to the exact solution can be obtained. A standard approach for determining approximate wave functions of bound states is the Rayleigh-Ritz variational method, used to great effect by James and Coolidge \cite{JamesCoolidge1933} and Ko{\l}os and Roothaan \cite{KolosRoothaan1960} in their pioneering calculations on the hydrogen molecule. Following these authors, we have taken the approximate wave function, \targetwave, to have the form

\begin{equation}\label{targetwave}
 \psi_{\mathrm{G}}=\sum_{v=1}^{L} c_{v}\varphi_{v},
\end{equation}
where

\begin{eqnarray}
\nonumber\varphi_{v}&=&\frac{1}{2\pi}\left(\lambda_{1}^{m_{v}}\lambda_{2}^{n_{v}}\mu_{1}^{j_{v}}\mu_{2}^{k_{v}}+\lambda_{1}^{n_{v}}\lambda_{2}^{m_{v}}\mu_{1}^{k_{v}}\mu_{2}^{j_{v}}\right)\\
\label{eq:targetbasis}&\times&s_{12}(\omega_{v}) \exp\left[-\delta\left(\lambda_{1}+\lambda_{2}\right)\right],
\end{eqnarray}
for prescribed basis states, $\{m_{v},n_{v},j_{v},k_{v},\omega_{v}\}$, comprising non-negative integers. The function $s_{12}(\omega_{v})$ has the same definition as used in (\ref{eq:s12}). Optimal values of the unknown constants $\{c_{v}\}$ are determined in the Rayleigh-Ritz method by minimizing the energy expectation of \targetwave.

The accuracy of \targetwave is typically measured in terms of the correlation energy of the molecule. This is the amount of the ground state energy, due to electron correlation, beyond that which is taken into account in a Hartree-Fock calculation \cite{BransdenJoachain2003}. The percentage, \pc, of the correlation energy accounted for by an approximate target wave function with ground state energy expectation, $E_{\mathrm{calc}}$, is

\begin{equation}\label{eq:pcorr}
 P_{\mathrm{C}}=\frac{E_{\mathrm{calc}}-E_{\mathrm{HF}}}{E_{\mathrm{ex}}-E_{\mathrm{HF}}} \times 100,
\end{equation}
where $E_{\mathrm{ex}}$ is the exact nonrelativistic ground state energy in the Born-Oppenheimer approximation \cite{BornOppenheimer1927} and $E_{\mathrm{HF}}$ is the Hartree-Fock energy.

We have carried out Kohn calculations using two different target wave functions, which for convenience we shall denote by \targA and \targB. The function \targA has $L=144$, with a basis set comprising 72 terms having $\omega_{v}=0$ and 72 terms having $\omega_{v}=2$. A value of $\delta=1.14$ was chosen for \targA to minimize its ground state energy expectation value, which accounted for $96.8 \%$ of the correlation energy of \hmol. \targA has the same general form as the 31-term function used in our earlier calculations \cite{Cooper2007}. The function \targB has a 145-term basis set of an identical form to that used for \targA, but for the inclusion of one \hyl term in $\rho_{12}$ for which $\omega_{v}=1$. The value of $\delta$ for \targB remained fixed at $\delta=1.14$, and the corresponding ground state energy accounted for $99.7 \%$ of the correlation energy of \hmol. Further details of the individual basis functions used are available from the corresponding author. The values of $E_{\mathrm{HF}}$ and $E_{\mathrm{ex}}$ used to determine \pc were taken respectively from the calculations of Jensen \cite{Jensen1999} and Wolniewicz \cite{Wolniewicz1994}.

The important role played by \hyl functions in $\rho_{12}$ in describing electronic correlations in the hydrogen molecule has long been known \cite{JamesCoolidge1933}. However, until very recently it was not feasible for us to carry out Kohn calculations with target functions of this form, due to difficulties in evaluating the corresponding integrals found in the Kohn equations. However, we have successfully made modifications to the computational framework used in our calculations so that target functions containing \hyl terms in $\rho_{12}$ can now be admitted. Earlier changes to this framework had already been made during our previous calculations \cite{Cooper2007}, where code designed originally for investigations of helium-antihydrogen scattering \cite{Armour2006} was adapted so that it could be applied to \ehmol scattering. Those initial modifications allowed for the evaluation of integrals containing terms in $\rho_{13}\rho_{23}/\rho_{12}$ by using a triple Neumann expansion \cite{ArmourHumberston1991,Plummer2007}. To carry out the calculations described here involving \targB, it was necessary to extend these modifications to allow for the evaluation of integrals containing factors of the form

\begin{equation}
G_{123}=F\left(\lambda_{\mathrm{3}}\right)\frac{\rho_{12}\rho_{23}}{\rho_{13}},
\end{equation}
and

\begin{equation}
G_{213}=F\left(\lambda_{\mathrm{3}}\right)\frac{\rho_{12}\rho_{13}}{\rho_{23}},
\end{equation}
where $F\left(\lambda_{\mathrm{3}}\right)$ can be either of the two open-channel functions (\ref{eq:SinOpenChannel},\ref{eq:CosOpenChannel}) representing the positron, or the function $\chi_{0}$.

\subsection{Optimization}\label{ss:opt}

In contrast to variational calculations of bound states, there is no energy minimization principle associated with scattering wave functions. As a result, there is no absolutely rigorous method available to optimize the nonlinear parameters, $\alpha$ and $\beta$, characterizing the rate of decay of the short-range correlation functions. Nevertheless, arguments for preferred choices of these parameters can be made. 

For atomic scattering, it has been shown \cite{Spruch1959} that, for a system where no bound state exists, the Kohn method gives an upper bound on the scattering length, $a$, where

\begin{equation}
 a=\lim_{k\rightarrow0}\left(-\frac{\tan\eta}{k}\right),
\end{equation}
and hence a lower bound on the exact phase shift, $\eta$, in the limit of zero positron energy. In the case of the Kohn variational method, obtaining bounds on scattering phase shifts is not generally possible at all incident energies considered, owing to the occurrence of the Schwartz anomalous behaviour at certain energies. However, an analysis of the method for potential scattering by Brownstein and McKinley \cite{Brownstein1968} showed that, away from these energies, the phase shift will be bounded, provided the trial functions are, in some sense, sufficiently accurate. 

In the case of the solution of scattering systems using a close-coupling expansion \cite{Bransden1983}, it has been shown that bounds exist on scattering phase shifts or eigenphase sums provided that all open channels are included in the expansion and that the open channel target states are exact \cite{Hahn1964,Gailitis1965}. If additional correlation functions are added to the expansion over target states, the bounds are still valid under certain conditions \cite{Hahn1964,Gailitis1965}, the extra terms acting as an optical potential for channels not explicitly included. 

Kohn calculations using exact target states have been carried out by Humberston and are described, for example, in \cite{ArmourHumberston1991}. He found that, at low energies, the variational approximation to the phase shift tended to increase monotonically as the flexibility of the trial wave function was improved by the inclusion of a greater number of short-range correlation functions. He concluded that it was reasonable to expect the variational approximation to converge upwards to the exact phase shift with the use of an increasingly flexible trial wave function. In the method of models it is assumed that the target wave function used in the calculation is an exact solution of a model problem, so that the Kohn scattering parameters converge to the exact values for the model. 

In the case of inexact target states, there are no known rigorous bounds on scattering parameters for general close-coupling calculations or for the Kohn method. The experience in both low energy elastic positron scattering \cite{VanReeth1995} using the Kohn method and close-coupling calculations of electron scattering (see, for example, \cite{Plummer2004,Zatsarinny2006}) is that the monotonic increase in phase shifts or eigenphase sums may continue well above the physical values if the description of scattering electronic correlation is made noticeably more intricate than the description of target electronic correlation. This highlights the need for determining whether the target states are sufficiently accurate to give reliable values for scattering parameters in a given calculation. 

Returning to our own implementation of the Kohn method, we see that, assuming our calculations are reliable in the sense that we have described here, we may regard \ps as an effective lower bound for $\eta$. Under these circumstances, values of the nonlinear parameters, $\alpha$ and $\beta$, can justifiably be chosen to maximize $\eta_{v}$. 

\section{Results and discussion}\label{s:results}

Our calculations were carried out for the lowest partial wave of $\Sigma_{\mathrm{g}}^{+}$ symmetry. A total of four different trial wave functions were used, corresponding to combinations of the two sets of correlation functions, $\Omega^{(1)}$ and $\Omega^{(2)}$, and the two inexact target functions, \targA and \targB. In a self-evident nomenclature, we will denote the four different trial wave functions by $\Psi_{\mathrm{t}}^{(1,\mathrm{A})}$, $\Psi_{\mathrm{t}}^{(2,\mathrm{A})}$, $\Psi_{\mathrm{t}}^{(1,\mathrm{B})}$ and $\Psi_{\mathrm{t}}^{(2,\mathrm{B})}$. Values of \ps and \zeff were determined for each trial wave function, for positron momenta in the range $k=0.01$ a.u. to $k=0.4$ a.u., corresponding to a maximum positron energy of $2.18$ eV. As noted in section \ref{s:introduction}, higher partial waves become significant in scattering processes above this energy.

Following our discussion in section \ref{ss:opt}, we selected values of $\alpha$ and $\beta$ approximately to maximize \ps. In principle, such maxima could be found, at least numerically, by straightforward iterative approaches. In practice, however, repeating our Kohn calculations for different values of $\alpha$ and $\beta$ is computationally very expensive; as we have discussed elsewhere \cite{Cooper2007}, each iteration necessitates the evaluation of a large number of integrals that can be obtained only numerically via a triple Neumann expansion. Our analyses were therefore restricted to a relatively small set of candidate values for the nonlinear parameters, namely, $\alpha \in \{0.2,0.3,\dots,0.9,1.0\}$ and $\beta \in \{0.2,0.3,\dots,1.4,1.5\}$. Kohn calculations were performed for each of the 126 combinations of $\alpha$ and $\beta$. Unless otherwise noted, all of the results presented here are from calculations carried out with $\alpha=0.3$ and $\beta=0.7$, which we found to maximize \ps for the trial wave function, $\Psi_{\mathrm{t}}^{(1,\mathrm{A})}$, at $k=0.04$. This value of $k$ is approximately equal to the average momentum of a Maxwellian distribution of positrons at $297$ K. It is convenient to consider the positron distribution at this temperature as it allows a direct comparison to be made of our results for \zeff with experimental data.

\subsection{Calculations involving \targA}\label{ss:psia}

\begin{figure}
 \centering
 \includegraphics{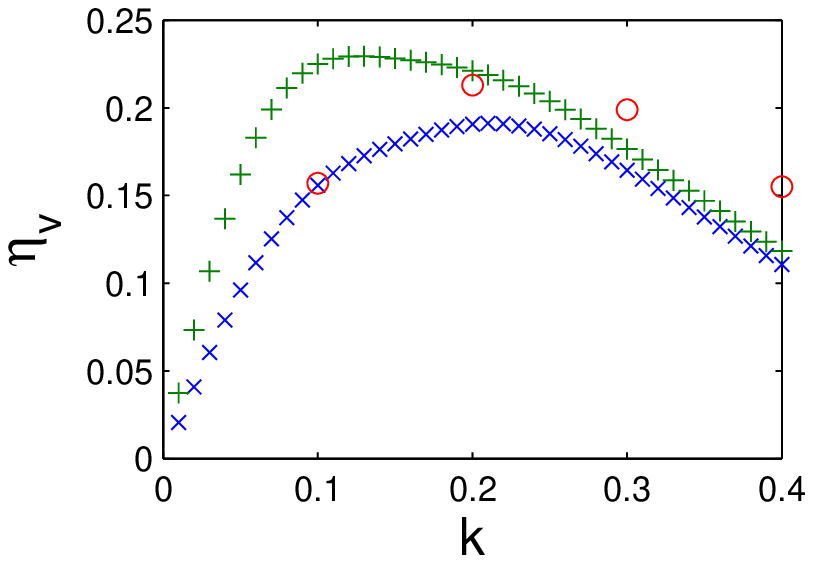}
 \caption{$\eta_{\mathrm{v}}\left(k\right)$ for [\textcolor{blue}{$\times$}]$\Psi_{\mathrm{t}}^{(1,\mathrm{A})}$, [\textcolor{myGreen}{$+$}]$\Psi_{\mathrm{t}}^{(2,\mathrm{A})}$ and [\textcolor{myRed}{\opencircle}], reported by Armour and Baker \cite{Armour1987}.}
 \label{fig:PSvK1}
\end{figure}

\begin{figure}
 \centering
 \includegraphics{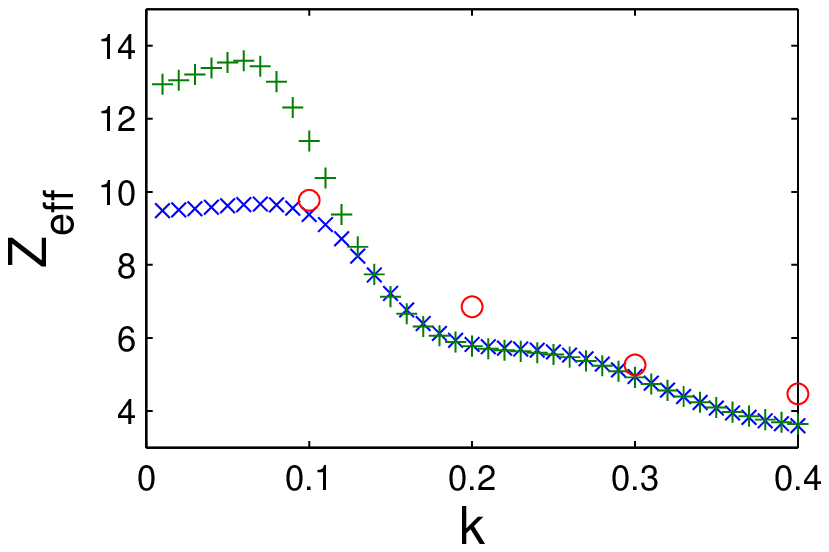}
 \caption{$Z_{\mathrm{eff}}\left(k\right)$ for [\textcolor{blue}{$\times$}]$\Psi_{\mathrm{t}}^{(1,\mathrm{A})}$, [\textcolor{myGreen}{$+$}]$\Psi_{\mathrm{t}}^{(2,\mathrm{A})}$ and [\textcolor{myRed}{\opencircle}], reported by Armour and Baker \cite{Armour1987}.}
 \label{fig:ZeffvK1}
\end{figure}

We consider first the two trial wave functions involving \targA, having $P_{\mathrm{C}}=96.8$. The dependence of \ps and \zeff on $k$ for $\Psi_{\mathrm{t}}^{(1,\mathrm{A})}$ and $\Psi_{\mathrm{t}}^{(2,\mathrm{A})}$ is shown in figures \ref{fig:PSvK1} and \ref{fig:ZeffvK1}. We have also included in these figures the values of \ps and \zeff found in table 2(e) of the account of Kohn calculations made by Armour and Baker \cite{Armour1987}. Those calculations used the method of models with a trial wave function containing $M=72$ correlation functions, including eight \hyl terms in the electron-positron distance.

The effect of including in $\Psi_{\mathrm{t}}^{(2,\mathrm{A})}$ the \hyl correlation functions in $\rho_{12}$ is clear. The calculated values of both \ps and \zeff for $\Psi_{\mathrm{t}}^{(2,\mathrm{A})}$ are significantly greater at low positron momenta than the corresponding values for $\Psi_{\mathrm{t}}^{(1,\mathrm{A})}$. The differences between the results for $\Psi_{\mathrm{t}}^{(1,\mathrm{A})}$ and $\Psi_{\mathrm{t}}^{(2,\mathrm{A})}$ become smaller at higher positron momenta. There is broad agreement between our results for \ps and those reported by Armour and Baker, although there is insufficient data available from those calculations to determine whether better agreement is observed for $\Psi_{\mathrm{t}}^{(1,\mathrm{A})}$ or $\Psi_{\mathrm{t}}^{(2,\mathrm{A})}$. Estimates, $\sigma_{\mathrm{v}}$, of the total scattering cross section could also be determined directly from values of \ps. However, there is a paucity of available experimental cross-section data at the very low positron momenta of most interest here, making any meaningful comparison with our results very difficult. 

There is good agreement between the estimates of \zeff for $\Psi_{\mathrm{t}}^{(1,\mathrm{A})}$, $\Psi_{\mathrm{t}}^{(2,\mathrm{A})}$ and the results of Armour and Baker, for $k\geq0.1$. Again, a lack of available data from those earlier calculations prevents a comparison below $k=0.1$, where the differences between the results for $\Psi_{\mathrm{t}}^{(1,\mathrm{A})}$ and $\Psi_{\mathrm{t}}^{(2,\mathrm{A})}$ are striking. We can, however, remark that the calculated value of \zeff for $\Psi_{\mathrm{t}}^{(2,\mathrm{A})}$ at $k=0.04$ is 13.4, in reasonable agreement with the accepted experimental value of $Z_{\mathrm{eff}}=14.61\pm0.14$ at $297$ K \cite{Laricchia1987}.

\begin{figure}
 \centering
 \includegraphics{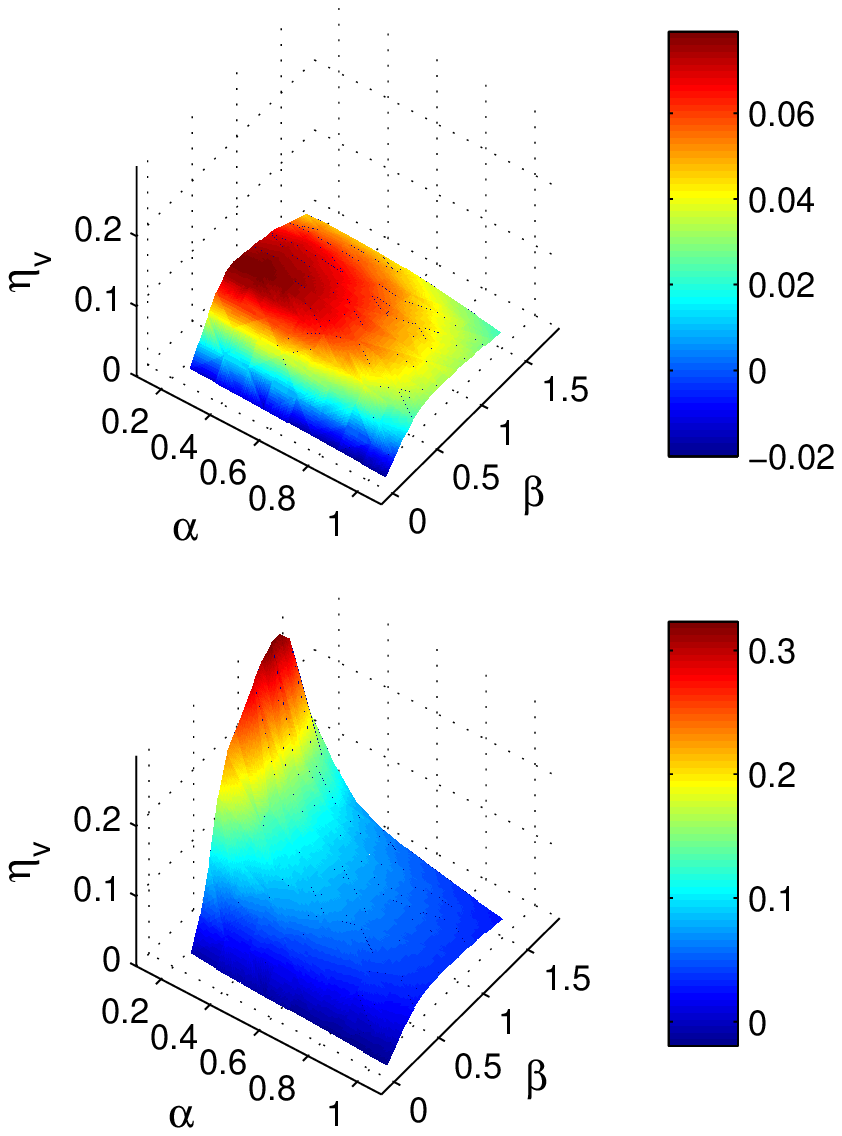}
 \caption{$\eta_{\mathrm{v}}\left(\alpha,\beta\right)$ at $k=0.04$, for $\Psi_{\mathrm{t}}^{(1,\mathrm{A})}$ (top) and $\Psi_{\mathrm{t}}^{(2,\mathrm{A})}$ (bottom).}
 \label{fig:PSsurf}
\end{figure}

\begin{figure}
 \centering
 \includegraphics{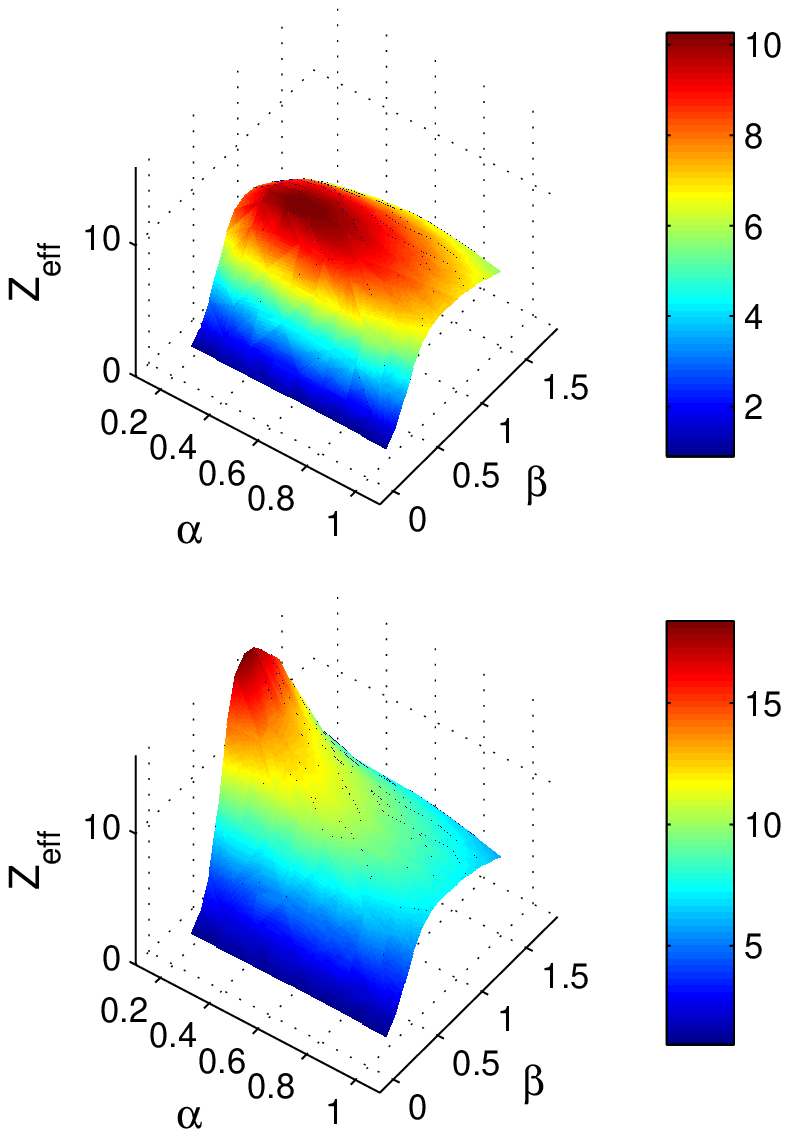}
 \caption{$Z_{\mathrm{eff}}\left(\alpha,\beta\right)$ at $k=0.04$, for $\Psi_{\mathrm{t}}^{(1,\mathrm{A})}$ (top) and $\Psi_{\mathrm{t}}^{(2,\mathrm{A})}$ (bottom).}
 \label{fig:Zeffsurf}
\end{figure}

The extent of the influence on \ps and \zeff of the \hyl functions in $\rho_{12}$ becomes even more pronounced if the values of the nonlinear parameters, $\alpha$ and $\beta$, are varied. Figures \ref{fig:PSsurf} and \ref{fig:Zeffsurf} illustrate the respective dependence of \ps and \zeff on $\alpha$ and $\beta$ at $k=0.04$, for $\Psi_{\mathrm{t}}^{(1,\mathrm{A})}$ and $\Psi_{\mathrm{t}}^{(2,\mathrm{A})}$. The effects of the \hyl functions in $\rho_{12}$ included in $\Psi_{\mathrm{t}}^{(2,\mathrm{A})}$ are most obvious for $\alpha<0.5$ and become more dramatic as the value of $\alpha$ decreases. Indeed, the values of \ps and \zeff for $\Psi_{\mathrm{t}}^{(2,\mathrm{A})}$ shown in figures \ref{fig:PSsurf} and \ref{fig:Zeffsurf} have not reached an obvious plateau with respect to further decreases in the value of $\alpha$. It seems plausible that these values would continue to increase with decreasing $\alpha$. 

In view of this, we think it necessary to examine the possibility that the observed effect is not genuine and instead arises from inaccuracies in the numerical evaluation of the integrals required to formulate the Kohn equations. These could occur because the short-range correlation functions become more diffuse as the value of $\alpha$ decreases, increasing the range of the configuration space of the positron over which the effects of the correlation functions are significant. To investigate this, we carried out a more detailed study of the Kohn calculations at $\left(\alpha,\beta\right)=\left(0.2,1.1\right)$ and $\left(\alpha,\beta\right)=\left(0.2,0.8\right)$, corresponding respectively to the maximum values of \ps and \zeff observed for $\Psi_{\mathrm{t}}^{(2,\mathrm{A})}$ in figures \ref{fig:PSsurf} and \ref{fig:Zeffsurf}. If the effects observed for $\Psi_{\mathrm{t}}^{(2,\mathrm{A})}$ are due to problems with convergence of integrals, increasing the range of integration in $\lambda$ should have a significant effect on the results for \ps and \zeff. However, when we increased the range of our integration in $\lambda$ by $50 \%$, the values of \ps at $\left(\alpha,\beta\right)=\left(0.2,1.1\right)$ and \zeff at $\left(\alpha,\beta\right)=\left(0.2,0.8\right)$  changed respectively by only $0.1 \%$ and $0.2 \%$ from the values shown for $\Psi_{\mathrm{t}}^{(2,\mathrm{A})}$ in figures \ref{fig:PSsurf} and \ref{fig:Zeffsurf}. This is clear evidence that the effects we have described are not due to errors in the numerical integration. 
 
\begin{figure}
 \centering
 \includegraphics{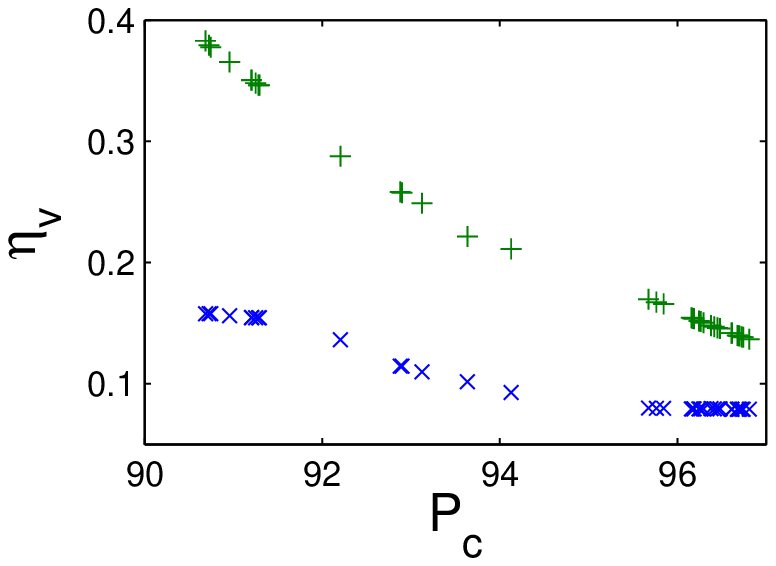}
 \caption{The dependence of \ps on \pc at $k=0.04$, for [\textcolor{blue}{$\times$}]$\Omega^{(1)}$ and [\textcolor{myGreen}{$+$}]$\Omega^{(2)}$. Basis functions have been removed successively from \targA.}
 \label{fig:pDep144PS}
\end{figure}

\begin{figure}
 \centering
 \includegraphics{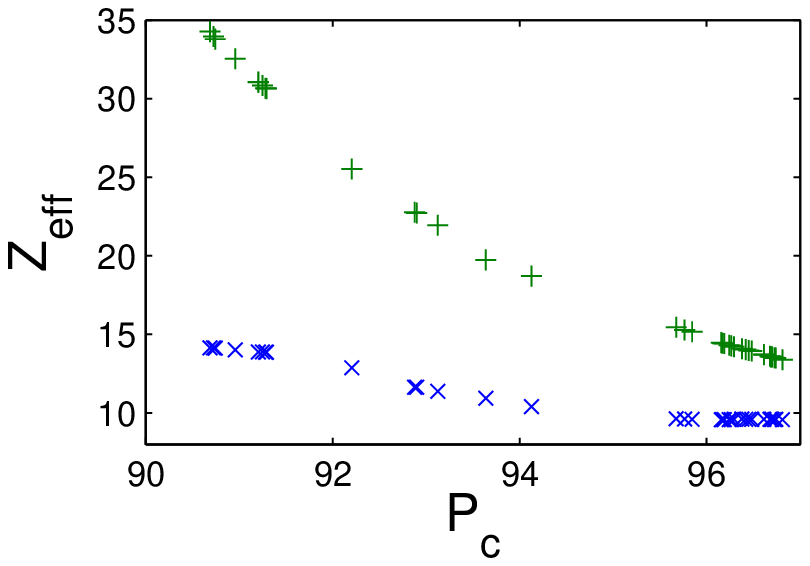}
 \caption{The dependence of \zeff on \pc at $k=0.04$, for [\textcolor{blue}{$\times$}]$\Omega^{(1)}$ and [\textcolor{myGreen}{$+$}]$\Omega^{(2)}$. Basis functions have been removed successively from \targA.}
 \label{fig:pDep144Zeff}
\end{figure}

Figures \ref{fig:PSvK1}--\ref{fig:Zeffsurf} indicate that the apparent importance of the \hyl correlation functions in $\rho_{12}$ is a general feature of the calculation at low positron momenta. This is unexpected, since functions of this type do not address explicitly the key difficulty of describing correlations in terms of the electron-positron separation. Following our discussion in section \ref{ss:opt}, it is conceivable that the observed behaviour is a result of inaccuracies in the calculation due to the use of the inexact target wave function, \targA, despite its taking into account of $96.8 \%$ of the correlation energy. This claim is consistent with the findings of Van Reeth and Humberston \cite{VanReeth1995}. In light of their conclusions, we investigated the sensitivity of our own Kohn calculations to changes in the accuracy of the target wave function. Basis functions were removed incrementally at random from \targA, creating a series of target wave functions of successively lower accuracies. After each removal, Kohn calculations were performed using the target wave function of reduced accuracy to determine values of \ps and \zeff, each time for two trial wave functions having the sets of correlation functions $\Omega^{(1)}$ and $\Omega^{(2)}$. A maximum of $70$ basis functions were removed from the original set of $144$ terms, at which point the target wave function accounted for $90.7 \%$ of the correlation energy of \hmol. The dependence of \ps and \zeff on the accuracy, \pc, of each target wave function is shown in figures \ref{fig:pDep144PS} and \ref{fig:pDep144Zeff} respectively, for $k=0.04$. 

The values of \ps and \zeff tend to decrease with increasing \pc, for both $\Omega^{(1)}$ and $\Omega^{(2)}$. However, there is a distinct convergence of \ps and \zeff for the results corresponding to $\Omega^{(1)}$ above $P_{\mathrm{C}}\sim96$, which is not evident for the results corresponding to $\Omega^{(2)}$. More generally, we have observed similar behaviour for analogous calculations performed at different values of $k$. We can reasonably conclude that Kohn calculations carried out using \targA are reliable when \hyl terms in $\rho_{12}$ are omitted from the correlation functions, but are unreliable when functions of this type are included. For trial wave functions containing the set of correlation functions, $\Omega^{(2)}$, we would expect any further increase in the accuracy of the target wave function to be accompanied by a noticeable drop in the values of \ps and \zeff. 

\subsection{Calculations involving \targB}\label{ss:psib}

\begin{figure}
 \centering
 \includegraphics{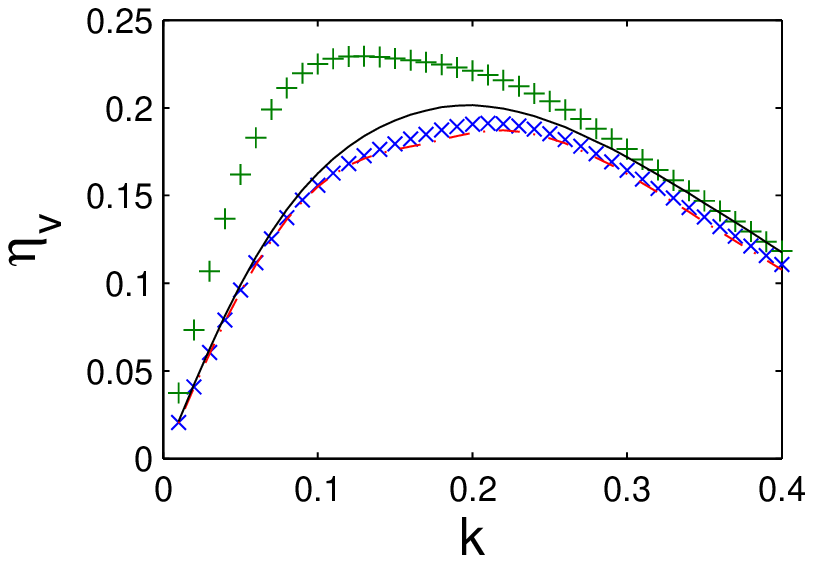}
 \caption{$\eta_{\mathrm{v}}\left(k\right)$ for [\textcolor{blue}{$\times$}]$\Psi_{\mathrm{t}}^{(1,\mathrm{A})}$, [\textcolor{myGreen}{$+$}]$\Psi_{\mathrm{t}}^{(2,\mathrm{A})}$, [\textcolor{myRed}{--$\cdot$--}]$\Psi_{\mathrm{t}}^{(1,\mathrm{B})}$ and [---]$\Psi_{\mathrm{t}}^{(2,\mathrm{B})}$.}
 \label{fig:PSvK2}
\end{figure}

\begin{figure}
 \centering
 \includegraphics{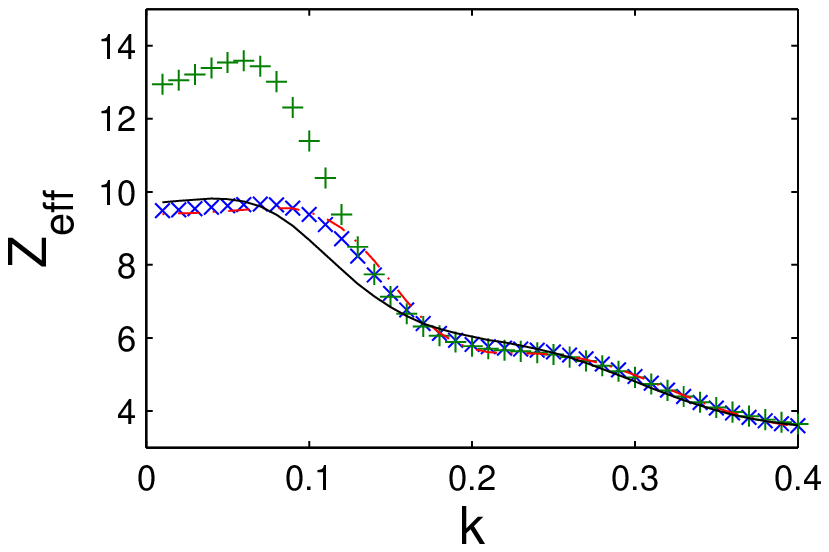}
 \caption{$Z_{\mathrm{eff}}\left(k\right)$ for [\textcolor{blue}{$\times$}]$\Psi_{\mathrm{t}}^{(1,\mathrm{A})}$, [\textcolor{myGreen}{$+$}]$\Psi_{\mathrm{t}}^{(2,\mathrm{A})}$, [\textcolor{myRed}{{\textbf{ --$\cdot$--}}}]$\Psi_{\mathrm{t}}^{(1,\mathrm{B})}$ and [---]$\Psi_{\mathrm{t}}^{(2,\mathrm{B})}$.}
 \label{fig:ZeffvK2}
\end{figure}

We consider now the two trial wave functions involving \targB, having $P_{\mathrm{C}}=99.7$. The corresponding Kohn calculations could be performed only when the modifications to the computational framework, described in section \ref{ss:thehydrogenmolecule}, had been implemented. The dependence of \ps and \zeff on $k$ for $\Psi_{\mathrm{t}}^{(1,\mathrm{B})}$ and $\Psi_{\mathrm{t}}^{(2,\mathrm{B})}$ is shown in figures \ref{fig:PSvK2} and \ref{fig:ZeffvK2}. We have also reproduced in these figures the values of \ps and \zeff determined earlier for $\Psi_{\mathrm{t}}^{(1,\mathrm{A})}$ and $\Psi_{\mathrm{t}}^{(2,\mathrm{A})}$.

A number of comments should be made about our results. Firstly, there is excellent agreement in the values of \ps and \zeff between calculations using $\Psi_{\mathrm{t}}^{(1,\mathrm{A})}$ and $\Psi_{\mathrm{t}}^{(1,\mathrm{B})}$, substantiating our earlier claim that \targA is a sufficiently accurate target wave function for the Kohn calculation using $\Psi_{\mathrm{t}}^{(1,\mathrm{A})}$ to have converged and hence to be considered reliable. Secondly, there are significant differences in the results for $\Psi_{\mathrm{t}}^{(2,\mathrm{A})}$ and $\Psi_{\mathrm{t}}^{(2,\mathrm{B})}$. The improvement in the accuracy of the target wave function has brought the results for $\Psi_{\mathrm{t}}^{(2,\mathrm{B})}$ broadly into line with those for $\Psi_{\mathrm{t}}^{(1,\mathrm{A})}$ and $\Psi_{\mathrm{t}}^{(1,\mathrm{B})}$. When the more accurate Kohn calculations using \targB are carried out, therefore, the effect of including \hyl correlation functions in $\rho_{12}$ is small.

\begin{figure}
 \centering
 \includegraphics{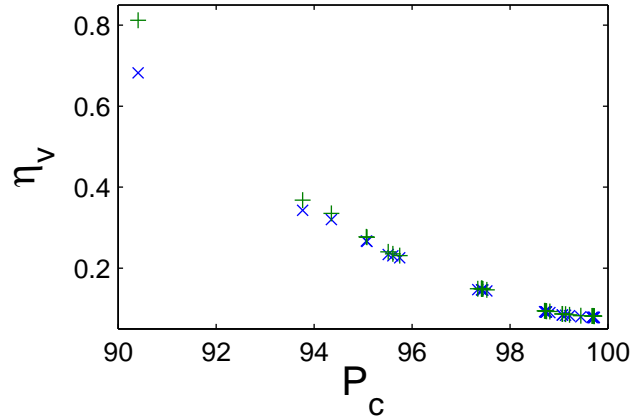}
 \caption{The dependence of \ps on \pc at $k=0.04$, for [\textcolor{blue}{$\times$}]$\Omega^{(1)}$ and [\textcolor{myGreen}{$+$}]$\Omega^{(2)}$. Basis functions have been removed successively from \targB.}
 \label{fig:pDep145PS}
\end{figure}

\begin{figure}
 \centering
 \includegraphics{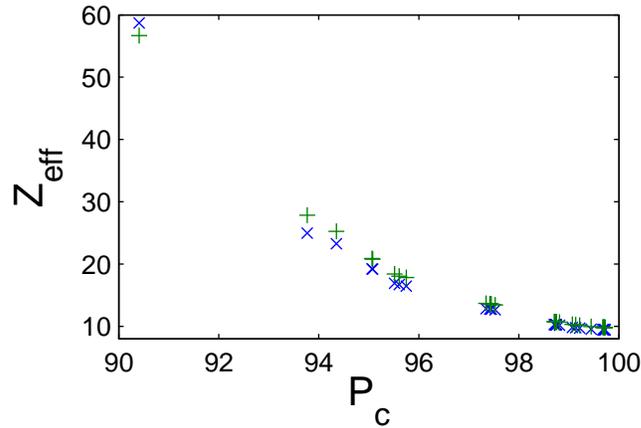}
 \caption{The dependence of \zeff on \pc at $k=0.04$, for [\textcolor{blue}{$\times$}]$\Omega^{(1)}$ and [\textcolor{myGreen}{$+$}]$\Omega^{(2)}$. Basis functions have been removed successively from \targB.}
 \label{fig:pDep145Zeff}
\end{figure}

It remains to be shown that \targB is a sufficiently accurate target wave function for calculations involving $\Psi_{\mathrm{t}}^{(2,\mathrm{B})}$ to be considered reliable. To do this, we again removed basis functions successively at random from the target wave function to reduce its accuracy. After each removal, Kohn calculations were performed using the target wave function of reduced accuracy to determine values of \ps and \zeff, each time for two trial wave functions having the sets of correlation functions $\Omega^{(1)}$ and $\Omega^{(2)}$. A maximum of $104$ basis functions were removed from the original set of $145$ terms comprising \targB, at which point the target wave function accounted for $90.4 \%$ of the correlation energy of \hmol. The first $70$ terms removed corresponded directly to the $70$ terms removed earlier from \targA. Thereafter, the remaining $34$ terms were removed successively at random, with the condition that the \hyl term in $\rho_{12}$ was not removed. The dependence of \ps and \zeff on the accuracy, \pc, of each target wave function is shown in figures \ref{fig:pDep145PS} and \ref{fig:pDep145Zeff} respectively, for $k=0.04$.

The convergence of \ps and \zeff with increasing \pc is not as obvious as it was in the previous calculations illustrated in figures \ref{fig:pDep144PS} and \ref{fig:pDep144Zeff}; however, the slopes of the curves in figures \ref{fig:pDep145PS} and \ref{fig:pDep145Zeff} are becoming noticeably flat in the upper limit of $P_{\mathrm{C}}$. We have already concluded that the Kohn calculations involving $\Psi_{\mathrm{t}}^{(1,\mathrm{A})}$, and hence those involving $\Psi_{\mathrm{t}}^{(1,\mathrm{B})}$, are reliable. Further, the behaviour of the curves corresponding to $\Omega^{(1)}$ and $\Omega^{(2)}$ is very similar in the limit of high \pc in both figures \ref{fig:pDep145PS} and \ref{fig:pDep145Zeff}. Inspection of these figures suggests that neither set of results would change significantly if the target wave function was again extended to account for most of the remaining $0.3 \%$ of the correlation energy. We may reasonably regard the calculations involving $\Psi_{\mathrm{t}}^{(2,\mathrm{B})}$ as reliable.

An interesting feature apparent from figures \ref{fig:pDep145PS} and \ref{fig:pDep145Zeff} is that the inclusion of the \hyl term in $\rho_{12}$ raises the threshold of convergence for trial wave functions containing $\Omega^{(1)}$. In these figures, the values of \ps and \zeff are still clearly declining at $P_{\mathrm{C}}=96$, at which value we have already concluded that Kohn calculations containing $\Omega^{(1)}$, using target wave functions without \hyl terms, have converged. The origin of this effect is not clear and will remain a subject of our investigations.

\section{Concluding remarks}

We have demonstrated that the reliability of Kohn calculations for \ehmol scattering can depend upon the flexibility of the correlation functions used in the trial wave function, relative to the flexibility and the accuracy of the approximate wave function representing the target. This dependence is most prominent at very low positron momenta. Our findings are similar to those reported by Van Reeth and Humberston for positron-helium scattering and highlight the need for rigorous testing of the accuracy of Kohn calculations whenever inexact target wave functions are used.

We have implemented a numerical method to test the stability of any given calculation to variations in the accuracy of the approximate target state. This has allowed us to distinguish between reliable and unreliable results and thus compensate for the lack of explicit bounds on the scattering phase shifts.

Having carried out the most accurate of our Kohn calculations, we have observed that the effect of including \hyl correlation functions in $\rho_{12}$ is to increase the calculated values of \ps only slightly. The changes in the values of \zeff are also small, so that there is still disagreement between our reported values and the established experimental result of $Z_{\mathrm{eff}}=14.61\pm0.14$ at $297$ K \cite{Laricchia1987}. This discrepancy is significant when compared to the results of other applications of the Kohn method for simpler systems. The best available calculations for atomic helium \cite{VanReeth1996}, for example, obtain a theoretical value of $Z_{\mathrm{eff}}=3.88\pm0.01$ at $293$ K. The experimental value is $3.94\pm0.02$ \cite{Coleman1975}. Our intention is to address the problems in our Kohn calculations for \hmol by improving the flexibility of the correlation functions still further to include, for example, terms linear in both $\rho_{12}$ and $\rho_{j3}$, $j\in\{1,2\}$, as well as terms describing virtual positronium formation. Consideration of virtual positronium has been shown \cite{VanReeth1998} to enhance significantly the calculated values of \zeff for positron scattering by atomic hydrogen near the positronium formation threshold. We hope that a similar increase in \zeff will be observed in our own calculations for molecular hydrogen if virtual positronium formation is taken into account. In any event, we will try to obtain converged results with as flexible a set of short-range correlation functions as possible.

\ack
We are grateful to John Humberston for valuable discussions. This work is supported by EPSRC (UK) grant EP/C548019/1.

\end{document}